\documentclass{article}
\usepackage{ijcai23}

\usepackage{times}
\usepackage{soul}
\usepackage{url}
\usepackage[hidelinks]{hyperref}
\usepackage[utf8]{inputenc}
\usepackage[small]{caption}
\usepackage{graphicx}
\usepackage{amsmath}
\usepackage{amssymb}
\usepackage{amsthm}
\usepackage{booktabs}
\usepackage{algorithm}
\usepackage[switch]{lineno}
\usepackage{bm}
\usepackage{algpseudocode}
\usepackage{xcolor}
\usepackage{natbib} 
\linenumbers

\nolinenumbers
\title{LM2D: Lyrics- and Music-Driven Dance Synthesis}

\author{
Wenjie Yin$^{1\ast}$\and
Xuejiao Zhao$^{1,2}$\thanks{Equal Contributions}\and
Yi Yu$^{3\dagger}$\and
Hang Yin$^4$\and
Danica Kragic$^1$\and
Mårten Björkman$^{1}$\thanks{Corresponding Authors}
\affiliations
$^1$KTH Royal Institute of Technology, 
$^2$Nanyang Technological University,\\ 
$^3$National Institute of Informatics, 
$^4$University of Copenhagen 
}

\begin{document}

\maketitle

\begin{abstract}
Dance typically involves professional choreography with complex movements that follow a musical rhythm and can also be influenced by lyrical content. The integration of lyrics in addition to the auditory dimension, enriches the foundational tone and makes motion generation more amenable to its semantic meanings. However, existing dance synthesis methods tend to model motions only conditioned on audio signals. In this work, we make two contributions to bridge this gap. First, we propose {LM2D}, a novel probabilistic architecture that incorporates a multimodal diffusion model with consistency distillation, designed to create dance conditioned on both music and lyrics in one diffusion generation step. Second, we introduce the first 3D dance-motion dataset that encompasses both music and lyrics, obtained with pose estimation technologies. We evaluate our model against music-only baseline models with objective metrics and human evaluations, including dancers and choreographers. The results demonstrate {LM2D} is able to produce realistic and diverse dance matching both lyrics and music. A video summary can be accessed at: \textit{\url{https://youtu.be/4XCgvYookvA}}. 
\end{abstract}

\section{Introduction}
\label{sec: introduction}
Dance is an engaging form of human expression that intertwines body movements with music, playing a crucial role in various cultures~\citep{lamothe2019dancing}. In the modern digital age, dance content enjoys enormous popularity on platforms like \textit{YouTube} and \textit{TikTok}, and even in video games such as \textit{Just Dance} and \textit{Dance Central}. 
However, creating dance, whether through traditional means or digitally, is a complex and challenging task. Professional dance involves expert choreography and extensive practice, often requires advanced motion capture technology for digitization. Consequently, the development of automated human motion generation technologies presents significant potential and possibilities across various digital platforms and the field of choreography. Such technology could pave the way for innovative collaborations between human creativity and artificial intelligence.

Recent breakthroughs in generative modeling, notably in the areas of normalizing flows~\citep{papamakarios2021normalizing} and diffusion models~\citep{ho2020denoising}, have significantly enhanced the capabilities of automated dance generation. Such advancements not only enrich the artistic dimension of choreography but also provide valuable insights for dance research~\citep{valle2021transflower,alexanderson2023listen,li2021ai,yin2023multimodal}. 

\begin{figure}[t]
  \centering
  \includegraphics[width=1\linewidth]{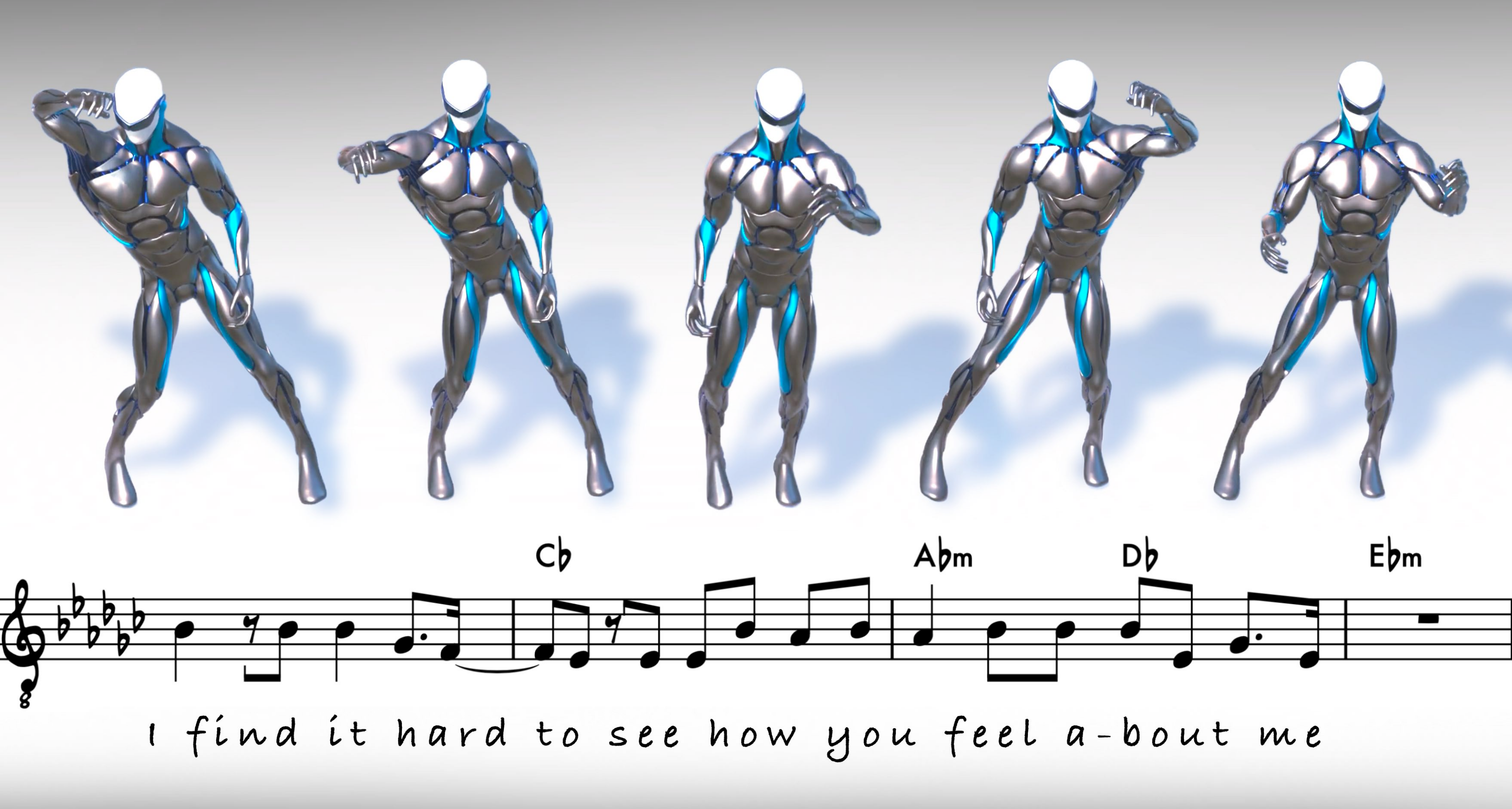}
  \caption{{LM2D}, a multimodal framework generates realistic and diverse dance movements conditioned on both lyrics and music. }
  \label{fig: teaser}
\end{figure}

Nevertheless, existing technologies in automated dance generation primarily focus on the interaction between music and dance movements, often neglecting the significant role of lyrics in choreography. While music-conditioned models can already produce realistic rhythmically-aligned dance movements, incorporating lyrics can add depth and enrich semantic meaning, as there is a notable connection between dance motion and song lyrics in styles like modern dance~\citep{powell2019modern}. Further exploration is needed to integrate both lyrics and music in dance synthesis. 
Additionally, the success of diffusion models and normalizing flows largely hinges on their iterative processes, which gradually build a sample from random noise. However, this step-by-step refinement, involving sequential steps of large neural network evaluations, results in slower sampling speeds compared to single-step methods like GANs or VAEs. This inefficiency poses a challenge for real-time applications, such as real-time dynamic choreography generation, emphasizing the need for efficient, single-step generation methods~\citep{song2023consistency,lu2022dpm}. Imagine a system capable of refreshing generated images at 5Hz in response to prompts editing in the context of image generation~\citep{sauer2023adversarial}. Such a real-time system in the context of dance choreography could facilitate instantly generating dance movements in response to changes in music and lyrics, enabling a highly interactive and responsive creative process. 

In response to these challenges, our study presents {LM2D}, a novel probabilistic architecture that combines a multimodal diffusion model with consistency distillation. As illustrated in Figure~\ref{fig: teaser}, this design aims to generate dance that is conditioned on both music and lyrics in a single diffusion step, addressing the limitations of existing models and advancing the field of automated dance generation. With consistency distillation, we could effectively distill a pre-trained diffusion model into a consistency model, which allows synthesizing dance with one evaluation step. 
Another major challenge in learning-based motion synthesis is the availability of large-scale datasets for 3D movement.
Existing dance datasets focus primarily on music and body motion, but lack lyric information. To bridge this gap, we apply pose estimation technologies~\citep{dong2020motion} on Just Dance videos to collect a new multimodal dataset that includes synchronized dance motion, music, and lyrics, providing a more comprehensive resource for research in this area. 
We evaluate our framework on the new dataset with the analysis focused on motion quality. A new metric is proposed to evaluate the match between motion and lyrics quantitatively. Additionally, we invite a group of human participants with extensive dancing and choreography experience to provide a subjective evaluation and insights from an expert perspective. These evaluations show that movements synthesized by {LM2D} are realistic and match both lyrics and music.
In summary, in this paper: 
\begin{itemize}
\item We contribute data-driven methods for lyrics- and music-driven dance synthesis. Our multimodal diffusion model with consistency distillation is able to create dance in a single diffusion step. 
\item We make a new dance dataset available. To the best of our knowledge, this is the first such dataset to contain synchronized lyrics, music, and high-quality 3D motion. 
\item We evaluate our new model both objectively and through a user study with skilled dancers and choreographers, focusing on the quality of motion and the alignments with lyrics and music. 
\end{itemize}

\section{Related Work}


\subsection{Data-Driven Dance Synthesis}
Dance synthesis, the problem of automatically creating realistic and natural human motions, is complex and challenging. 
Early research employed motion retrieval methods, creating choreography by transitioning between pre-existing motion clips~\citep{fan2011example,lee2013music,fukayama2015music}. These methods, which generate motion by selection, often led to unnatural transitions and had limited variability. 
With the advent of deep learning, \citep{ye2020choreonet,chen2021choreomaster} have integrated deep learning techniques with motion graphs to produce higher-quality choreography. 
Subsequent research trained on large datasets and explored various modeling approaches, including generative adversarial networks (GANs), recurrent neural networks (RNNs), transformers, and normalizing flows~\citep{fan2022bi,li2022danceformer,li2021ai,yin2023multimodal,siyao2022bailando,valle2021transflower}.
Recent breakthroughs with diffusion models~\citep{ho2020denoising} have further advanced this field. 
\citep{alexanderson2023listen} pioneered the use of diffusion models with Conformer~\citep{zhang2022music} for generating dance from music. 
EDGE~\citep{tseng2023edge} and Magic~\citep{li2022magic} utilized Transformer-based diffusion models. 
However, these methods primarily focus on music-driven synthesis, largely neglecting the impact of lyrics, which usually contain rich semantic information. 
A recent study by \citep{deichler2023diffusion} has ventured into using joint text and audio representation for human gesture generation. 
In our research, we introduce \textit{LM2D}, a model that generates dance movements conditioned by both lyrics and music using diffusion models. 
Additionally, diffusion models typically exhibit slower sampling speeds compared to single-step methods such as GANs or VAEs. To overcome this limitation, we incorporate consistency distillation to accelerate the inference process to a single diffusion step. 

\subsection{Dance Datasets}
3D dance datasets are essential for data-driven dance synthesis, which requires professional experience and usually contains multi-modal information. 
Motion capture techniques are widely adopted for collecting 3D motion data. The first notable 3D dance dataset was released by~\citep{alemi2017groovenet} with synchronized music. Following this, \citep{zhuang2022music2dance} collected {Music2Dance}, a synchronized music and motion capture dataset. 
Recently, \citep{valle2021transflower} introduced three new datasets, the PMSD, Syrto, and ShaderMotion VR dance datasets, each focusing on different dance styles. Among these, the ShaderMotion VR dance dataset incorporates VR technology for collection. 
\citep{alexanderson2023listen} combined various data sources to create a high-quality dataset. 
DanceFormer~\citep{li2022danceformer} presented the PhantomDance dataset, a unique collection created by professional animators, while ChoreoMaster~\citep{chen2021choreomaster} combined motion capture and anime resources in their dataset.
Advancements in 3D reconstruction have enabled the conversion of 2D videos into 3D dance data, which allows for the acquisition of 3D skeletal data from a vast amount of videos, at a lower cost compared to motion capture or animation techniques.
\citep{li2021ai} utilized the AIST dance dataset~\citep{tsuchida2019aist} to obtain 3D dance motion, employing multi-view 3D pose estimation techniques, and \citep{li2020learning} extracted 3D pose sequences with synchronized audios using the VideoPose3d~\citep{pavllo20193d}. 
\citep{le2023music} and \citep{wang2022groupdancer} introduce group dance datasets for more challenging group choreography. 
However, to the best of our knowledge, there is no dataset available that synchronizes music, lyrics, and body motion. Existing human motion datasets typically include either music with motion or text with body gestures. To bridge this gap, we have constructed a new dataset using 3D pose estimation, captureing dance sequences synchronized with both music and lyrics from JustDance video games.

\section{Method}

\begin{figure}[t!]
\centering
  \includegraphics[width=1\linewidth]{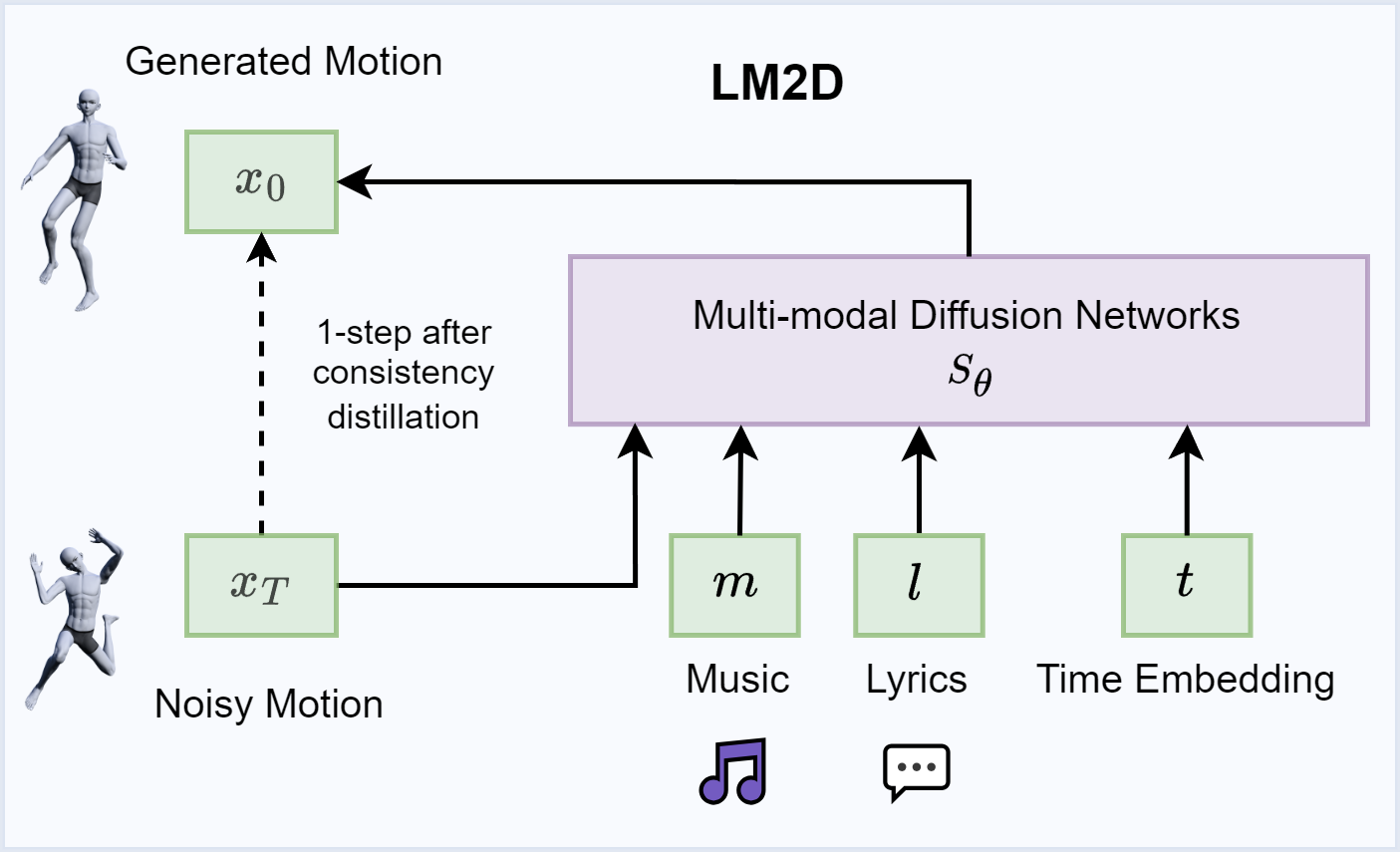}   
  \caption{
  Overview of the LM2D framework. LM2D learns to denoise dance sequences from time $t=T$ to $t=0$, condition on music and lyrics in one step with consistency distillation.  
  }
  \label{fig: model}
\end{figure}


Our approach uses a multi-modal neural network to produce a series of human poses, $\bm{x}$, based on a corresponding sequence of music features $\bm{m}$ and lyrics features $\bm{l}$, as illustrated in Figure \ref{fig: model}. In this section, we delve into the mathematical properties of diffusion models and the architecture for generating motion driven by both lyrics and music. Subsequently, we introduce the concept of consistency distillation for the purpose of facilitating one-step generation. 

\subsection{Multi-modal Diffusion Framework} 

To tackle the problem described above, we follow the theory of continuous-time diffusion models~\citep{song2020score}. 
Diffusion models produce samples by gradually transforming data into noise through Gaussian disturbances and then generating samples from noise through a series of denoising steps.
In the continuous-time diffusion model, the diffusion process $\{\bm{x}(t), t \in [0, 1] \}$ can be defined as a forward Stochastic Differential Equation (SDE)~\citep{song2020score}:
\begin{equation}
    d\bm{x}_t = \bm{f}(\bm{x}_t, {t}) d{t} + g({t})d\bm{w}_t,
\label{eqn: forward_sde}
\end{equation}
where $\bm{w}_t$ represents the standard Wiener process operates in reverse-time, $\bm{f}(\bm{x}_t, t)$ denotes the drift term, and $g(t)$ is the scalar diffusion coefficient. 
The reverse-time SDE specifies the reverse process of the above forward process as follows:
\begin{equation}
d\bm{x}_t = \left[ \bm{f}(\bm{x}_t, {t})  - g(t)^{2} \nabla_{\bm{x}_t} \log p_{\it{t}}(\bm{x}_t)\right] \, d{t} + g({t}) \, d\bm{w}_t,
\end{equation}
where $\nabla_{\bm{x}_t}\log p_t(\bm{x}_t)$ represents the score function associated with the data distribution perturbed by noise at time $t$. 
For the reverse-time SDE, there exists a specific ordinary differential equation (ODE) known as the Probability Flow ODE (PF-ODE)~\citep{song2019generative,song2020score}. The solutions of the PF-ODE sampled at time $t$ are distributed according to $p_{\it{t}}(\bm{x}_t)$:
\begin{equation}
\frac{d\bm{x}_t}{d{t}} =\bm{f}(\bm{x}_t, {t})  - \frac{1}{2}g(t)^{2} \nabla_{\bm{x}_t} \log p_{\it{t}}(\bm{x}_t),
\end{equation}
In the training phase, a score model $\bm{s}_{\theta}(\bm{x}_t,t)$ is employed to approximate the term $-\nabla_{\bm{x}_t} \log p_{\it{t}}(\bm{x}_t)$ through score matching~\citep{song2019generative,song2020score}, which leads to the following ODE for sampling:
\begin{equation}
\frac{d\bm{x}_t}{d{t}} =\bm{f}(\bm{x}_t, {t})  + \frac{g(t)^{2}}{2}\bm{s}_{\theta}(\bm{x}_t,t). 
\end{equation}
We initialize the PF-ODE by sampling $\bm{z}_T \sim \mathcal{N}(\bm{0}, \bm{I})$ and then solve it in reverse time using a numerical ODE solver.
The obtained $\hat{\bm{x}}_0$ can be regarded as an estimated sample of the original data distribution. The solver is typically stopped at $t=\epsilon$ to avoid numerical instability issues, where $\epsilon$ is a predetermined small positive value.

Our diffusion framework is developed based on the EDGE architecture~\citep{tseng2023edge}, designed for producing human motion sequences given conditional signals. 
This framework employs a transformer-based network with cross-attention mechanisms~\citep{li2021ai} that accepts music $\bm{m}$, lyrics $\bm{l}$ features, latents $z_t$ that follow a forward noising process to estimate $\hat{\bm{x}}_\theta$. 
Subsequently, it synthesizes motion sequences that align with these features, as illustrated in Figure \ref{fig: model}.
The objective function is simplified as:
\begin{equation}
\mathcal{L}_{rec}=\mathbb{E}_{\bm{x}, t}\left\|\bm{x} - \hat{\bm{x}}_\theta(\bm{z}_t, {t}, \bm{m}, \bm{l}) \right\|^2_2.
\end{equation}
In addition to the above reconstruction loss, we also adopt geometric losses as~\citep{tang2022real,tevet2022human,tseng2023edge} to improve physical realism with joint positions and velocities.
\begin{equation}
\mathcal{L}_{pos} = \frac{1}{N} \sum_{i=1}^{N} \| FK(\bm{x}^i - FK(\hat{\bm{x}}^i) \|_2^2,
\end{equation}
\begin{equation}
\mathcal{L}_{vel} = \frac{1}{N-1} \sum_{i=1}^{N-1} \| (\bm{x}^{i+1} - \bm{x}^i) - (\hat{\bm{x}}^{i+1} - \hat{\bm{x}}^{i}) \|_2^2,
\end{equation}
where $FK(\cdot)$ is the forward kinematic function, $N$ is the number of frames in the synthesized sequences. 
The overall training loss is a weighted sum of the reconstruction loss and geometric losses:
\begin{equation}
\mathcal{L} = \mathcal{L}_{rec}+\lambda_{pos}\mathcal{L}_{pos}+\lambda_{vel}\mathcal{L}_{vel}.
\end{equation}

\subsection{Consistency Distillation} 

\begin{figure}[t!]
\centering
  \includegraphics[width=1\linewidth]{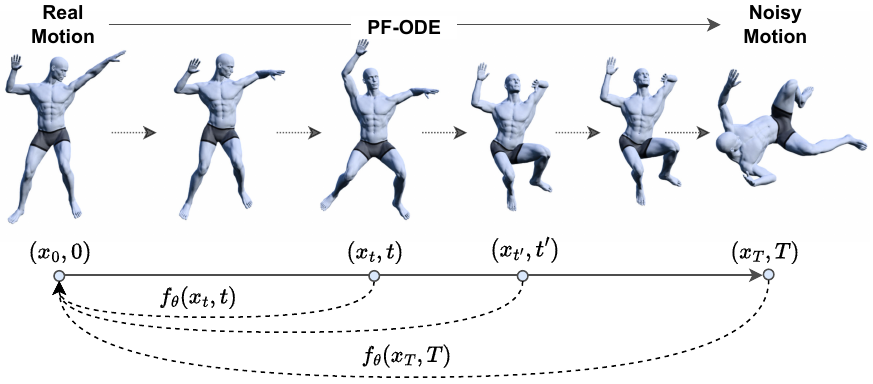}   
  \caption{
  Overview of the consistency models. Given a PF-ODE that smoothly converts real human motion to noisy motion, we learn to map any points on the trajectory to its origin point. 
  }
  \label{fig: cd}
\end{figure}

The performance of diffusion models is constrained by their slow sampling process. As the ODE solvers for sampling necessitate multiple iterations of the score model $s_\theta$, ranging from dozens to thousands of steps, leading to significant computational demands. 

To accelerate the diffusion sampling, the consistency models~\citep{song2023consistency} are introduced. The consistency model is a new family of generative models capable of executing single-step generation while preserving generation quality. 
These models can be trained either by distilling pre-existing diffusion models or as independent generative models. In our research, we apply the method of consistency distillation (CD) to achieve one-step generation. For the sake of brevity, we have excluded the conditional music and lyrics features in this section.
The key idea of CD focuses on learning the function that maps every point on a PF-ODE trajectory to its origin point.
More formally, the consistency function is defined as $\bm{f}_\theta:\left(\bm{x}_t, t\right) \rightarrow \bm{x}_\epsilon$, and the consistency function should satisfy the self-consistency property:
\begin{equation}
\bm{f}_\theta(\bm{x}_t, t) = \bm{f}_\theta(\bm{x}_{t'}, t'), \forall t, t' \in [\epsilon, T],
\end{equation}
which can be parameterized using skip connections: 
\begin{equation}
\bm{f}_\theta(\bm{x}_t, t) = c_{\text{skip}}(t)\bm{x} + c_{\text{out}}(t)\bm{S}_\theta(\bm{x}, t),
\end{equation}
where $c_{\text{skip}}(t)$and $c_{\text{out}}(t)$ are differentiable functions with $c_{\text{skip}}(\epsilon)=1$ and $c_{\text{out}}(\epsilon)=0$, and $\bm{S}_\theta(\bm{x}, t)$ is the neural network. 
The consistency distillation loss is defined as
\begin{equation}
\mathcal{L}(\theta, \theta^{-}; \Phi) = \mathbb{E}_{\bm{x},t_n} \left[ d\left( \bm{f}_\theta(\bm{x}_{t_{n+1}}, t_{n+1}), \bm{f}_{\theta^{-}}(\bm{\hat{x}^{\phi}}_{t_n}, t_n) \right) \right],
\end{equation}
here the time horizon $[\epsilon, T]$ is discretized into $n=N-1$ sub-intervals. $d(\cdot,\cdot)$ is the $\ell_2$ metric for measuring the distance between samples, $\theta^{-}$ is the parameters of the target model, updated with the exponential moving average (EMA) of the online model $\theta$, i.e., $\theta^{-} \leftarrow \mu\theta^{-}+(1-\mu)\theta$, $\mu$ is the decay rate. 
$\Phi$ is the ODE solver applied to the PF-ODE, $\bm{\hat{x}^{\phi}}_{t}$ is a one-step estimation of $\bm{x_{t}}$ from $\bm{x_{t+1}}$ as:
\begin{equation}
\bm{\hat{x}^{\phi}}_{t_n} \leftarrow \bm{x}_{t_{n+1}} + (t_n-t_{n+1})\Phi(\bm{x}_{t_{n+1}}, t_{n+1};\phi).
\end{equation}
We employ consistent distillation to learn consistency models from the pre-trained multi-modal diffusion models, achieving one-step dance generation given music and lyrics features. The algorithm of consistency distillation for the pre-trained multi-modal LM2D is depicted in Algorithm \ref{alg: cd}. For more details, please refer to~\citep{song2023consistency}.  

\begin{algorithm}[t]
\caption{Consistency Distillation (CD) for LM2D}\label{alg: cd}
\begin{algorithmic}
\State \textbf{Input:} pre-trained model parameter $\theta$, ODE solver $\Phi$, learning rate $\eta$, and training data pairs of motion $x$, music $m$, and lyrics $l$. 
$\theta^{-} \leftarrow \theta$
\State \textbf{Repeat:} \\
    \quad $\bm{\hat{x}^{\phi}}_{t_n} \leftarrow \bm{x}_{t_{n+1}} + (t_n-t_{n+1})\Phi(\bm{x}_{t_{n+1}}, m, l, t_{n+1};\phi)$
    \quad
    \begin{align*}
    \mathcal{L}(\theta, \theta^{-};  &\, \Phi) \leftarrow\\
    &\,d(\bm{f}_\theta(\bm{x}_{t_{n+1}}, m,l, t_{n+1}), \bm{f}_{\theta^{-}}(\bm{\hat{x}^{\phi}}_{t_n}, m, l, t_n))
    \end{align*}
    \quad $\theta \leftarrow \theta - \eta \nabla_\theta\mathcal{L}(\theta, \theta^{-};  \Phi)$\\
    \quad $\theta^{-} \leftarrow \mu\theta^{-}+(1-\mu)\theta$
\State \textbf{Until} convergence 
\end{algorithmic}
\end{algorithm}

\section{Dataset}
\subsection{Data Collection} 
We create a new dance dataset due to the lack of datasets simultaneously containing dance motion, music, and lyrics. 
The proposed 3D motion dataset is derived from existing Just Dance videos by Ubisoft, a motion-based rhythm dancing game with annual releases, and has been a well-known classic in video games. Just Dance features a diverse range of dance styles including pop, hip-hop, Latin, classical, and electronic dance music. The game involves players replicating the movements of an on-screen dancer. Utilizing EasyMocap~\citep{dong2020motion}, we extracted 3D human motion data in terms of SMPL parameters from these videos. This method allowed us to achieve high-fidelity body estimations at 60 fps. Additionally, we compiled a multimodal dataset by sourcing music and lyrics. The lyrics for each song were manually gathered and synchronized with the corresponding musical timeline. In summary, this dataset encompasses 4.6 hours of 3D dance motion in 1867 sequences, accompanied by music and lyrics. We will publicly release the data. 

\subsection{Data Preparation} 
We represent dance as sequences of poses using the 24-joint SMPL format~\citep{loper2015smpl}, using a 6-DOF rotation representation~\citep{zhou2019continuity}, resulting in a 147-dimensional feature.
We follow the audio representation as in ~\citep{valle2021transflower,tsuchida2019aist} that combines spectrogram features with beat-related features. The music features are extracted by Librosa~\citep{mcfee2015librosa}, yielding a 35-dimensional feature. Specifically, this encompasses a combination of 20-dimensional Mel-frequency cepstral coefficients (MFCC), 12-dimensional chroma, and 1-dimensional one-hot peaks and beats.
Lyrics were processed and embedded into a pre-trained BERT embedding~\citep{devlin2018bert}, resulting in a 768-dimensional feature.

\subsection{Data Validity} 
The task of synthesizing dance from both music and lyrics could benefit from several considerations:
1) {Semantic Influence}: There is a connection between the semantics of motion and lyrics. For example, the word \textit{No} is naturally associated with gestures like shaking one's head or hands; 
2) {Emotional Influence}: Lyrics can enhance the emotion the music conveys. For instance, phrases like \textit{Break my heart }might be interpreted through quieter and sadder movements;   
3) {Rhythmic Pattern Influence}: The patterns in rhythms and lyrics are often related. Repetitive lyrical patterns such as \textit{Oh oh oh} or \textit{Ma-ma-ma} might suggest a sequence of recurring dance motions;
4) {Music-lyrics Influence}: Different segments of a song may have identical music but distinct lyrics. Incorporating lyrics into the choreography can help avoid producing repetitive movements in similar musical segments. 
While these considerations might not always apply, they are not uncommon in various choreographies. 

\section{Experiments}
\label{sec: exp}

\subsection{Experimental Setting} 

To demonstrate the capabilities of our proposed LM2D, we compare it to baseline models and ground truth on the introduced new dataset. 
To evaluate the impact of lyrics in dance motion synthesis, we compare our model with EDGE, which is only conditioned on music information. Additionally, we compare the trained diffusion models with the one-step model with consistency distillation to explore the performance of consistency distillation on motion generation. 
To evaluate the performance, we quantitatively assessed (1) FID scores and diversity; (2) beat alignment scores; and (3) our proposed semantic match metric. The objective evaluation results are shown in Table \ref{tab: obj} and discussed in Section \ref{sec: obj}. 
Qualitative human evaluations are also important in the evaluation of generative models as well. We performed a user study in the form of an online survey to evaluate human-perceived quality in terms of (1) motion naturalness; (2) motion-music alignment; and (3) motion-lyrics matching. The subjective evaluation results are discussed in Section \ref{sec: sub}. 

\subsection{Objective Evaluation} 
\label{sec: obj}

\begin{table}[t]
\resizebox{\linewidth}{!}{
\begin{tabular}{@{}ccccccc@{}}
\toprule
Method       & $FID_k\downarrow$ & $FID_g\downarrow$ & $Div_k\rightarrow$ & $Div_g\rightarrow$ & BA$\uparrow$  & \quad SM$\uparrow\quad$    \\ \midrule
\quad Ground Truth & -                 & -                 & 11.34                     & 7.47                      & 0.24          & 0.85          \\ \midrule
EDGE         & 12.81             & 9.25              & \textbf{12.33 }                    & 7.18                      & \textbf{0.25}          & 0.81          \\ \midrule
EDGE(cd)     & \textbf{11.40}            & \textbf{8.74}     & 12.43                     & 6.83                      & 0.22          & 0.80          \\ \midrule
LM2D         & 12.35             & 9.76              & 12.37            & \textbf{7.53}             & \textbf{0.25} & \textbf{0.83} \\ \midrule
LM2D(cd)     & 12.1    & 10.35             & 12.58                     & 7.37                      & 0.24          & 0.83          \\ \bottomrule
\end{tabular}}
\caption{Quantitative objective evaluation: FID score and diversity metrics based on geometric and kinetic features are computed. Besides, beat alignment score (BA) and semantic matching score (SM) are calculated for evaluating alignments with motions. }
\label{tab: obj}
\end{table}

\paragraph{FID Scores.} 
In our study, we initially considered the Fréchet Inception Distance (FID) score on geometric and kinetic features. The FID score is a widely recognized objective metric for evaluating generative models and has been extensively used in previous research~\citep{li2021ai,li2020learning}. This metric quantifies the divergence between the distributions of real and synthetic data. However, recent work, ~\citep{tseng2023edge} has pointed out that the FID score may not be entirely reliable for tasks involving dance generation. They observed that the FID results can contradict human evaluations, possibly due to subjective judgments. In our experiments, as shown in Table \ref{tab: obj}, the EDGE model showed improved results after consistency distillation, but this improvement was not reflected in the human evaluations. Thus, although we include the results of FID score, the utility of FID for assessment can be questioned.

\paragraph{Diversity Metrics.} 
Diversity metrics in our study were calculated based on the distributional spread of geometric and kinetic features. In line with the methodologies employed in prior research~\citep{li2021ai, tseng2023edge}, our model aims to match this metric with those of the ground truth distribution. As indicated in Table~\ref{tab: obj}, the diversity metrics among the various models are quite similar, suggesting that the distributions generated by each model are closely aligned with the real distribution. Specifically, the kinetic features of EDGE are the closest to the ground truth, while the geometric features of LM2D most closely match the ground truth. However, it was observed that after applying consistency distillation to accelerate the model, the diversity gap between the synthetic and real distributions increased.

\paragraph{Beat Alignment Scores.} 
Our experiments included an evaluation of how well our generated dances aligned with the beat of the music, as in previous work by \citep{siyao2022bailando} and \citep{tseng2023edge}. The findings indicated that both EDGE and LM2D models achieved comparable levels of performance in terms of beat alignment scores. However, it was also observed that the application of consistency distillation led to a decrease in this score. 

\paragraph{Semantic Matching Scores} 

Our task extends beyond musical elements to include lyrical content, necessitating an evaluation of both movement-music alignment and movement-lyric correspondence. To measure the semantic matching between movements and lyrics, we utilized the pre-trained BERT model~\citep{devlin2018bert} to obtain BERT embeddings for the lyrics. For the semantic features of the movements, we followed the implementation of MotionBert~\citep{zhu2023motionbert}, using our multimodal dataset to train a motion encoder that captures motion embeddings. During the training process, we kept the parameters of the pre-trained BERT model fixed and trained the motion encoder to better align the features of lyrics and movements in our dataset. We evaluated the semantic matching by calculating the cosine similarity between the motion and lyric embeddings. Observing the experimental results, the LM2D model, which incorporates lyric information during training, achieved higher Semantic matching scores compared to the EDGE model, which does not include lyric information. This demonstrates the effectiveness of incorporating lyrics into the training process for enhanced semantic matching in dance generation. 

\begin{figure}[t!]
\centering
  \includegraphics[width=1\linewidth]{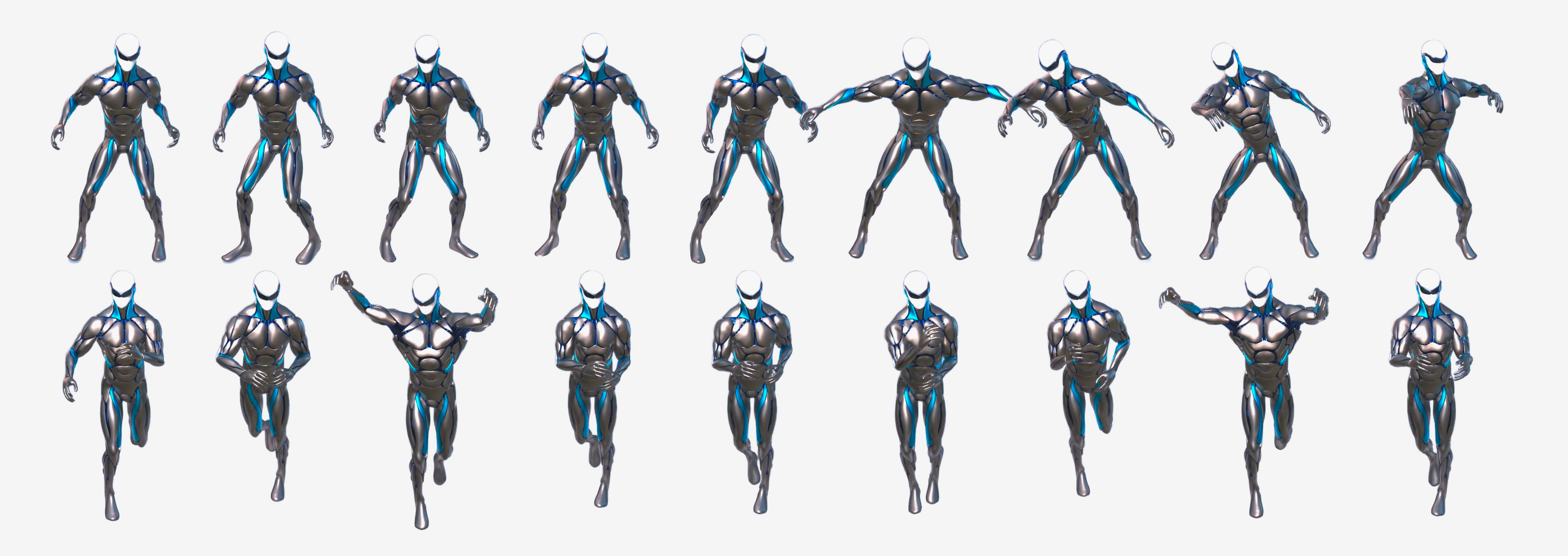}   
  \caption{LM2D Example: Two dance sequences are generated from the \textbf{same music} but with \textbf{different lyrics}. 
  }
  \label{fig: ex1}
\end{figure}

\begin{figure}[t!]
\centering
  \includegraphics[width=1\linewidth]{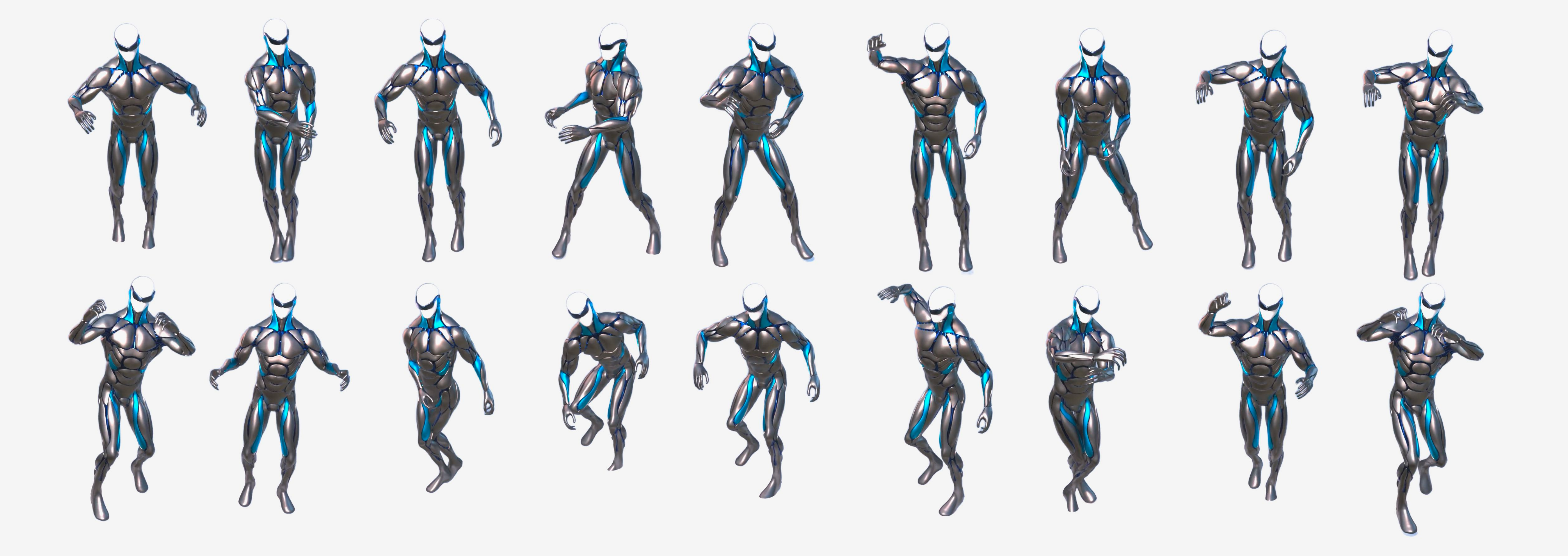}   
  \caption{
  LM2D Example: Two dance sequences are generated from the \textbf{same lyrics} but with \textbf{different music}. 
  }
  \label{fig: ex2}
\end{figure}

Figure \ref{fig: ex1} and Figure \ref{fig: ex2} present examples of synthesized motion clips from LM2D, which demonstrate the combined influence of lyrics and music on dance motion synthesis. 
Figure \ref{fig: ex1} showcases examples where two dance sequences were generated from the same musical piece, \textit{``Think About Things,''} but with differing lyrics. The upper sequence was generated with the lyrics \textit{``I wouldn't stop movin' if I could Break it down!''} while the lower sequence used \textit{``Just call me For delivery. They know me. I got sushi."} Despite the identical musical input, the distinct lyrics led to different movement patterns in each sequence. Additionally, the movements are, to some extent, semantically aligned with the lyrics.
In Figure \ref{fig: ex2}, we present dance sequences generated using the same lyrics but different music segments. The lyrics used were \textit{``Every time I'm around there's a hype! Touchdown and the crowd get's hyped."} The upper music segment is from \textit{``Bad Guy,"} while the lower one is from \textit{``Hype."} It's noticeable that, due to the consistency of the lyrics, there are similarities in the movement patterns between the sequences, particularly in the positions of the limbs and torso. However, the differences in music necessitate that the movements also align with the musical beats. This approach allows us to generate more diverse and meaningful dance movements.

\subsection{Human Evaluations.} 
\label{sec: sub}

To gain a deeper understanding of our method, we perform human evaluations alongside the objective evaluation. In this study, participants were asked to assess three key aspects: motion naturalness, motion-music alignment, and motion-lyrics matching. 
Our study recruited 30 participants with dance experience, including training, performing, teaching, or even choreography. The average dance experience among these participants was 6.72 years. Notably, 7 of the participants (23\% of the group) had experience in choreography.
We conducted an online survey to collect feedback from these participants to evaluate the task of synthesizing dance driven by lyrics and music. We blended 6-second video clips representing the ground truth and created motion sequences. Participants were shown these dance video clips, followed by a series of questions. 
Participants were instructed to rate their level of agreement with the statements presented in the questions, using a 1-5 Likert scale, where ``1" indicates strong disagreement, and ``5" indicates strong agreement.
To mitigate potential order effects, the sequence in which the dance clips were presented was randomized and balanced across participants.

In the first part of the survey, we focused on evaluating two aspects: the naturalness of the motion and the alignment between motion and music. The dance clips provided for evaluation were generated using one of several methods: LM2D, EDGE, LM2D-cd, EDGE-cd, or they were sourced from Ground Truth (GT). Participants had the option to view these clips multiple times, enabling a thorough assessment before responding to the following two questions:
\begin{itemize}
\item{\textbf{Motion Naturalness}}: 
\textit{To what extent do you agree with the following statement? --- The movements are natural and of good quality.} 
\item{\textbf{Motion-Music Alignment}}: 
\textit{To what extent do you agree with the following statement? --- The movements match well with the beat of the music. } 
\end{itemize}
In the second part of the survey, our evaluation centered on the matching between motion and lyrics. For this part, each dance clip was either generated using the LM2D method or taken from Ground Truth. Participants need to respond to the following question after viewing the clips:
\begin{itemize}
\item{\textbf{Motion-Lyrics Matching}}: 
\textit{To what extent do you agree with the following statement? --- The movements match well with the lyrics.}
\end{itemize}

We performed a statistical analysis of the subjective responses from the user study to support our findings and evaluated whether our proposed method could be further enhanced.

\begin{figure}[t!]
\centering
  \includegraphics[width=1\linewidth]{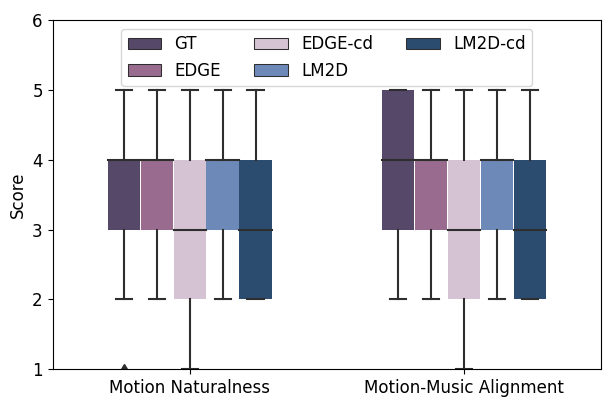}   
  \caption{
  Subjective evaluation results in motion naturalness and motion-music alignment.
  }
  \label{fig: u1}
\end{figure}

\begin{figure}[t!]
\centering
  \includegraphics[width=1\linewidth]{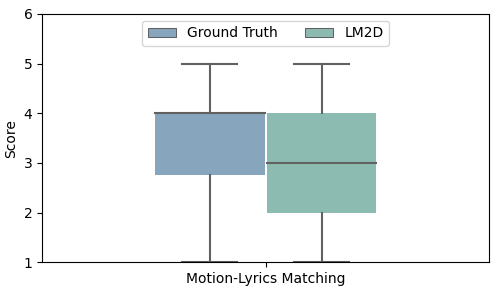}   
  \caption{
  Subjective evaluation results in motion-lyrics matching.
  }
  \label{fig: u2}
\end{figure}

Figure \ref{fig: u1} illustrates the evaluation of motion naturalness and motion-music alignment. From the figure, we observe that participants considered the performance of EDGE and LM2D to be similar to the Ground Truth (GT). However, models enhanced with consistency distillation (cd) showed a slight decline in performance, aligning with objective findings. We employed Analysis of Variance (ANOVA) to test statistically significant differences among the groups.

For motion naturalness, ANOVA results showed no significant difference among GT, EDGE, and LM2D $(F = 0.0191, p = 0.981)$. The same was observed for motion-music alignment $(F = 0.1134, p = 0.8929)$. These results demonstrate that our model, even after incorporating lyric content, maintains the motion naturalness and the alignment with the music's rhythm while incorporating semantic content with the movements. 
In contrast, for GT, EDGE-cd, and LM2D-cd, significant differences were observed for motion naturalness $(F = 5.2529, p = 0.0061)$ and motion-music alignment $(F = 12.1948, p = 1.0923 \times 10^{-5})$, as per ANOVA. This suggests that while consistency distillation enhances efficiency, it adversely affects performance, a finding echoed by human evaluations and indicating a need for further improvement. Figure \ref{fig: u2} focuses on comparing the LM2D model with ground truth regarding motion-lyrics matching. 
Dance experts favored the ground truth. We applied the two one-sided tests to assess if the evaluations of the two systems are statistically equivalent. The tests for equivalence did not provide sufficient evidence to confirm statistical equivalence between the LM2D model and ground truth $(p = 0.799, \delta=0.05)$. This highlights the challenge of achieving parity with ground truth in terms of motion-lyrics matching.

This study further integrates open-ended questions to gather in-depth feedback and identify areas for future enhancements. Experts with choreography experience suggested the inclusion of criteria to evaluate choreography more comprehensively. They noted that while certain movements are a good match and appear natural, they are relatively simple. Additionally, the evaluation of subtle, smaller movements presented challenges. For example, movements involving the neck are relatively subtle, yet they appear somewhat stiff.
This limitation is partly due to the dance dataset being sourced from Just Dance, a rhythm game developed for the general public that does not encompass highly complex or challenging dance sequences. 
Regarding the matching of motion to lyrics, a common observation was that some videos did not align well with the lyrics. This issue is largely attributed to the data, where the focus of the movements was primarily to follow the musical beats rather than to match the lyrics precisely. However, it's important to recognize that dance is not a literal translation of words like sign language. Effective choreography should consider the overall harmony of rhythm and lyrics rather than mechanically translating lyrics into movements.
Additionally, choreography experts have expressed the need for real-time modification of generated movements based on lyrics. Our work further investigates consistency distillation to achieve applicability in this area. However, upon comparison, the performance of distillation techniques in dance generation can still be further improved.

\section{Ethical and Societal Discussion}
This work presents a new dataset and a system for creating dances, aiming to bring new ideas and improvements to the field of dance. It facilitates the generation of dance influenced by lyrics and music, offering valuable tools for artists and researchers. This work could also be advantageous to fields like video gaming and animation.
Nevertheless, the automation of choreographic processes raises concerns about originality and creativity that it may blur the lines of ownership in creative endeavors.

\section{Conclusions and Future Work}
In this study, we introduce LM2D, a multimodal diffusion-based model for generating realistic dance sequences conditioned on both lyrics and music. A novel dance dataset featuring a combination of lyrics, music, and 3D motion data was collected. We enhanced synthesis efficiency by employing consistency distillation for one-step generation. Our extensive quantitative and qualitative assessments demonstrate the effectiveness and superior capabilities of the proposed method. In summary, this research paves the way for creating complex choreographies that are synchronized with musical rhythms and lyric semantics. 

Looking ahead, our future endeavors include integrating Large Language Models (LLMs) to deepen the understanding of lyrics. LLMs have demonstrated impressive capabilities in a wide range of NLP tasks, also evidenced by recent advancements in understanding, planning, and generating motion with LLMs~\citep{zhang2023t2m,zhou2023avatargpt}. The community should leverage these powerful LLMs to push the boundaries of dance motion synthesis, potentially synthesizing multiple modalities simultaneously, such as dance motion and audio that includes sung lyrics. In addition, future work will focus on achieving accelerated output generation while maintaining the quality of the results.

\section{Acknowledgments}
This research received partial support from the National Institute of Informatics (NII) in Tokyo. This work has been supported by the European Research Council (BIRD-884807) and H2020 EnTimeMent (no. 824160). 
This work was partially supported by the Wallenberg Al, Autonomous Systems and Software Program (WASP) funded by the Knut and Alice Wallenberg Foundation. 
This work also benefited from access to the HPC resources provided by the Swedish National Infrastructure for Computing (SNIC), partially funded by the Swedish Research Council through grant agreement no. 2018-05973. 

\bibliographystyle{named}
\bibliography{ijcai23}

\end{document}